\documentclass{article}

\usepackage{arxiv}
\usepackage{subcaption}
\usepackage{amsmath}
\usepackage[utf8]{inputenc} 
\usepackage[T1]{fontenc}    
\usepackage{hyperref}       
\usepackage{url}            
\usepackage{booktabs}       
\usepackage{amsfonts}       
\usepackage{nicefrac}       
\usepackage{microtype}      
\usepackage{lipsum}		
\usepackage{graphicx}

\title{Field-Embedded Factorization Machines for Click-Through Rate Prediction}

\author{Harshit Pande\thanks{This work was done when the author was employed at Adobe} \\
	Adobe\\
	\texttt{pandeconscious@gmail.com} \\
}

\date{}

\renewcommand{\shorttitle}{Field-Embedded Factorization Machines for Click-Through Rate Prediction}

\hypersetup{
	pdftitle={Field-Embedded Factorization Machines for Click-Through Rate Prediction},
	pdfsubject={cs.LG},
	pdfauthor={Harshit Pande},
	pdfkeywords={factorization machines, recommender systems, feature interactions, deep learning CTR, matrix embeddings},
}

\begin{document}
	\maketitle

\begin{abstract}
  Click-through rate (CTR) prediction models are common in many online applications such as digital advertising and recommender systems. Field-Aware Factorization Machine (FFM) and Field-weighted Factorization Machine (FwFM) are state-of-the-art among the shallow models for CTR prediction. Recently, many deep learning-based models have also been proposed. Among deeper models, DeepFM, xDeepFM, AutoInt+, and FiBiNet are state-of-the-art models. The deeper models combine a core architectural component, which learns explicit feature interactions, with a deep neural network (DNN) component. We propose a novel shallow Field-Embedded Factorization Machine (FEFM) and its deep counterpart Deep Field-Embedded Factorization Machine (DeepFEFM). FEFM learns symmetric matrix embeddings for each field pair along with the usual single vector embeddings for each feature. FEFM has significantly lower model complexity than FFM and roughly the same complexity as FwFM. FEFM also has insightful mathematical properties about important fields and field interactions. DeepFEFM combines the FEFM interaction vectors learned by the FEFM component with a DNN and is thus able to learn higher order interactions. We conducted comprehensive experiments over a wide range of hyperparameters on two large publicly available real-world datasets. When comparing test AUC and log loss, the results show that FEFM and DeepFEFM outperform the existing state-of-the-art shallow and deep models for CTR prediction tasks. We have made the code of FEFM and DeepFEFM available in the DeepCTR library(\url{https://github.com/shenweichen/DeepCTR}).
  
\end{abstract}




\section{Introduction}
\label{sec:intro}
The click-through rate (CTR) prediction models play an important role in the rapidly growing multi-billion dollar online advertising industry~\cite{revenuereport2018, mcmahan2013ad}. In a CTR prediction system, the main task is to predict the probability of click on an item shown to a user. Displaying items or advertisements of desired preference to a user not only boosts revenue but also enhances user satisfaction. The CTR data is usually multi-categorical and modeling interaction between features across fields leads to improved CTR prediction performance.  Because of the high number of categories in such data, the number of unique features can be very high, often in the order of millions. Multiple studies have shown a link between lift in the CTR prediction performance and improvement in online CTR \& increase in revenue~\cite{deepfm2017, widedeep20016}.  The studies show that a small lift in offline CTR performance leads to significant increase in revenue by improving the online CTR.


Shallow factorization machine-based models~\cite{logistic2007, fm2010, ffm2016, fwfm2018}, Bayesian methods \cite{bayesiangraepel2010}, tensor factorization methods \cite{tensorkoren2009} and tree-based methods \cite{treemethods2014} have been used to model CTR prediction tasks. Given the  breakthrough  in deep neural networks~\cite{lecun2015deep}, incorporation of  deep learning  for CTR prediction tasks is an active area of research~\cite{widedeep20016, fnn2016deep, pnn2016deep, deepcrosswang2017, NFMhe2017, deepfm2017, xiao2017attentional, DINzhou2018, fpennliu2018, song2019autoint, xdeepfm2019, fibinethuang2019}. We discuss both the existing shallow and deep models in further detail in Section~\ref{sec:prelims-related-work}. A feature may behave differently depending on the field of its interacting feature. We refer to this property as field specificity. The shallow models Field-Aware Factorization Machine (FFM) \cite{ffm2016} and Field-weighted Factorization Machine (FwFM) \cite{fwfm2018} use this property for improved performance. In deep learning models, FiBiNet~\cite{fibinethuang2019} and Field-aware Probabilistic Embedding Neural Network (FPENN)~\cite{fpennliu2018} also use this property for higher performance. Neural networks that use layers designed to capture the properties of their respective data domains are successful. The success of convolutional layers ~\cite{lecun1998gradient} for images and Long shot-term memory(LSTM) layers ~\cite{hochreiter1997long} for text is partly due to their ability to exhibit the inductive bias needed for these respective domains. Field specificity is one such property of the CTR tasks that we also use to design the core layers of our proposed models. Another interesting property observed in the existing literature is that at most 3-order or 4-order interactions are needed to give the best CTR prediction performance, while further higher-order interactions lead to degradation in performance. The number of DNN layers is usually less than or equal to 3 and a similar depth is needed for the core layers proposed by different methods such as Deep Cross Network DCN~\cite{deepcrosswang2017}, Compressed Interaction Network (CIN) in xDeepFM~\cite{xdeepfm2019},   Squeeze-Excitation network (SENET) module in FibiNet~\cite{fibinethuang2019}, and self-attention modules in AutoInt~\cite{song2019autoint}. With these observations in mind, we also designed an architecture of our deep learning models to capture at most 3 or 4-order interactions, while designing the layers with inductive bias needed to model CTR prediction tasks.


Here we list the key contributions of this paper:

\begin{itemize}
	\item We first introduce shallow Field Embedded Factorization Machine (FEFM), a novel factorization machine variant,  that outperforms the existing state-of-the-art shallow models such as FM, FFM and FwFM on publicly available CTR datasets. We introduce symmetric field pair matrix embeddings to capture field specificity for feature interactions. FEFM has significantly lower model complexity (in terms of number of parameters) than FFM and similar model complexity as FM and FwFM.
	
	\item The eigen values of the matrix embeddings of FEFM provide an interpretable way of modeling field pair interactions. We define such field interaction strengths in terms of these eigen values and demonstrate that the fields appearing at the top of the list of strongest interactions are intuitively expected to do so.
	
	\item We extend our proposed shallow FEFM to a deeper variant called Deep Field Embedded Factorization Machine (DeepFEFM). DeepFEFM uses field pair matrix embeddings to generate FEFM interaction embeddings, which in combination with feature vector embeddings via skip concatenation layers generate higher order interactions. On doing extensive experiments on publicly available CTR datasets, we show that DeepFEFM consistently outperforms other state-of-the-art deep models such as DeepFM, xDeepFM, AutoInt, and FiBiNet.
	
	\item We compare our models and existing state-of-the-art models for multiple embedding dimensions to evaluate not only an overall best performing model but also to study the effect of embedding dimensions on different models. This is in contrast with most of the existing deep learning work, which compares models for only one embedding dimension. Interesting and novel outcomes arise as a result of this study.
\end{itemize}

The rest of the paper is organized as follows. In Section ~\ref{sec:prelims-related-work} we provide the mathematical notations, preliminaries and a discussion on the related work . In Section ~\ref{sec:proposed-models}, we introduce our proposed models in detail. In Section ~\ref{sec:experimentsresults}, we present the experimental setup and the results on public data sets. The conclusions are presented in Section ~\ref{sec:conclusion}.

\section{Preliminaries and Related work}
\label{sec:prelims-related-work}

Suppose there are $n$ unique fields and $m$ unique features across all the fields in a CTR prediction system where $F(i)$ is the field to which the $i^{th}$ feature belongs. In practice $n \ll m$. Suppose there are $N$ points in data with $d^{th}$ data point defined as $(y^{(d)}, x^{(d)})$, where $y^{(d)} \in \{1, -1\}$  ($1$ and $-1$ indicate click and no-click respectively) and $x^{(d)} \in \{0,1\}^m$ is a multi-hot vector indicating $x_i^{(d)} = 1$ if feature i is active and $x_i^{(d)} = 0$ if feature i is inactive. Irrespective of the modeling technique, it is common to have the click probability  modeled as a logistic function $\sigma(\phi) = \frac{1}{1+\exp(-\phi)}$. The optimization problem given in Equation \ref{eq:optimization} forms the basis of many well-known CTR prediction methods. 

\begin{equation}
\label{eq:optimization}
min_{\theta} \: \lambda \frac{\|\theta\|_2^2}{2} + \sum_{d=1}^{N} log(1 + exp(-y^{(d)}\phi(\theta, x^{(d)})))
\end{equation}

where $\lambda$ is the regularization strength. Different type of parameters may have different regularization strengths. For simplicity we show a single $\lambda$. The difference among methods lies in how each of the methods model the logit function function $\phi$ and what parameters $\theta$ are used in the models. A logistic regression (LR) model ~\cite{logistic2007} serves as a baseline model and its logit function is shown in Equation~\ref{eq:lm-modeling}. 

\begin{equation}
\label{eq:lm-modeling}
\phi(\theta, x) = \phi_{LR}(w, x) = w_0 + \sum_{i=1}^{m} w_ix_i 
\end{equation}
The linear parameter $w$ is often also used in factorization machine-variant models (as a linear component) along with the specific parameters of the respective models.

Factorization Machine (FM) and variants~\cite{fm2010, ffm2016, fwfm2018} learn vector embeddings for each feature in the data and model the feature interactions in different ways. Equation \ref{eq:fm-modeling-general} serves as a base equation for multiple shallow variants of FM. In practice the summation over large $m$ is not needed as only the number of features in the order of much smaller $n$ are active for one data point.

\begin{equation}
\label{eq:fm-modeling-general}
\begin{split}
\phi(\theta, x) = \phi_{FMvariant}((v,w), x) = w_0 + \sum_{i=1}^{m} w_ix_i + \sum_{i=1}^{m}\sum_{j=i+1}^{m}Interaction(i,j)
\end{split}
\end{equation}
$Interaction(i,j)$ models the interaction between the $i^{th}$ feature and the $j^{th}$ feature. The modeling of this interaction term varies for different FM-variants. Suppose the parameter $v_i$ is the vector embedding for the $i^{th}$ feature and $v_j$ for the $j^{th}$ feature. For FM~\cite{fm2010}, the interaction $Interaction(i,j)$ is modeled as $v_i^Tv_jx_ix_j$. Field-Aware Factorization Machine (FFM) \cite{ffm2016} outperform FM by incorporating field specificity. For every feature, FFM learns multiple embeddings, one separate embedding per field. The$Interaction(i,j)$ for FFM is $v_{i,F(j)}^Tv_{j,F(i)}x_ix_j$. Here the parameter $v_{i,F(j)}$ is the vector embedding for the $i^{th}$ feature when it interacts with a feature from the field $F(j)$. Similarly, the parameter $v_{j,F(i)}$ is the vector embedding for the $j^{th}$ feature when it interacts with a feature from the field $F(i)$. A large number of parameters, leading to a high memory footprint, causes a major hindrance to the use of FFM in real-world production systems. Field-weighted Factorization Machine (FwFM) \cite{fwfm2018} extends FM by adding another parameter which models field specificity as the strength of field pair interactions. $Interaction(i,j)$ for FwFM is $v_i^Tv_jr_{F(i),F(j)}x_ix_j$, where the scalar parameter $r_{F(i),F(j)}$ represents interaction strength between fields $F(i)$ and $F(j)$. FwFM uses only a scalar to model field specificity. There is potential of further improvement by using higher dimensional matrix embeddings for modeling field specificity, which we use in our proposed models.

Now, we discuss the related work in deep learning. The most unifying underlying property of these deep learning methods is the combination of features generated by their respective core layers with feed forward deep neural networks (DNN). The major difference lies in the design of the core layers. Neural factorization machine (NFM)~\cite{NFMhe2017}, Deep Factorization Machine~\cite{deepfm2017}, and Factorization-machine supported Neural Networks (FNN)~\cite{fnn2016deep} use the combination of a factorization machine variant along with DNNs. Our proposed DeepFEFM shares some similarities with DeepFM and NFM. The DNN part of DeepFM takes input only the input feature embeddings, but we additionally also input FEFM generated interaction embeddings. Our ablation studies in Section~\ref{sec:experimentsresults} show that these FEFM interaction embeddings are key to the lift in performance achieved by DeepFEFM. NFM generates embeddings fed to a DNN via  bi-interaction pooling layer, which are generated as a Hadamard product between feature embeddings. Product-based Neural Networks (PNN)~\cite{pnn2016deep}, Deep Cross Network (DCN)~\cite{deepcrosswang2017}  Deep Interest Network (DNN)~\cite{DINzhou2018}, extreme deep factorization machine (xDeepFM)~\cite{xdeepfm2019}  use custom-designed core layers along with DNNs. They also combine their core layers with DNNs but their way of generating the intermediate layers is quite different from our method. Wide and Deep models~\cite{widedeep20016}, Attentional Factorization Machine (AFM), AutoInt~\cite{song2019autoint}, Feature Importance and Bilinear feature Interaction Network (FibiNet)~\cite{fibinethuang2019} have their core architecture inspired from successful deep learning models from other domains. FPENN~\cite{fpennliu2018} estimates the probability distribution of the field-aware embedding as compared to FFM which is a single point estimation. However, for deep learning, they have only compared their method with DeepFM. FiBiNet shares similarity to our models as it also uses pairwise field matrices in their Bilinear-Interaction layer, but they use a combination of inner and outer product, which is different from our FEFM layer used for generating interaction embeddings.

\section{Proposed Models}
\label{sec:proposed-models}
We propose a shallow model Field-Embedded Factorization Machines (FEFM) and its extension for deep learning, which we call Deep Field-Embedded Factorization Machine (DeepFEFM)

\subsection{Field-Embedded Factorization Machines}
\label{sec:fefms}
For each field pair, Field-Embedded Factorization Machines (FEFM) introduces symmetric matrix embeddings along with the usual feature vector embeddings that are present in FM. Like FM, $v_i$ is the vector embedding of the $i^{th}$ feature. However, unlike FFM, FEFM doesn't explicitly learn field-specific feature embeddings. The learnable symmetric matrix $W_{F(i),F(j)}$ is the  embedding for the field pair $F(i)$ and $F(j)$. The interaction between the $i^{th}$ feature and the $j^{th}$ feature is mediated through $W_{F(i),F(j)}$. This modeling is shown in Equation \ref{eq:fefm-modeling}. 

\begin{equation}
\label{eq:fefm-modeling}
\begin{split}
\phi(\theta, x) = \phi_{FEFM}((w, v, W), x) = w_0 + \sum_{i=1}^{m} w_ix_i  + \sum_{i=1}^{m}\sum_{j=i+1}^{m} v_i^TW_{F(i),F(j)}v_jx_ix_j
\end{split}
\end{equation}
where $W_{F(i),F(j)}$ is a $k \times k$ symmetric matrix ($k$ is the dimension of the feature vector embedding space containing feature vectors $v_i$ and $v_j$).

The symmetric property of the learnable matrix $W_{F(i),F(j)}$ is ensured by reparameterizing $W_{F(i),F(j)}$  as $U_{F(i),F(j)} + U^T_{F(i),F(j)}$, where $U^T_{F(i),F(j)}$ is the transpose of the learnable matrix $U_{F(i),F(j)}$.  In section ~\ref{subsec:interaction-strength}, we discuss how the symmetric property of $W_{F(i),F(j)}$ helps it in modeling field pair interaction strengths. Note that $W_{F(i),F(j)}$ can also be interpreted as a vector transformation matrix which transforms a feature embedding when interacting with a specific field. This is advantageous over FFM, which instead explicitly learns field specific-feature embeddings, causing huge number of parameters and a tendency of overfitting. Since both the number of fields ($n$) and the embedding dimensions ($k$) are much smaller than number of features ($m$), FEFM uses less number of parameters while modeling similar field-specificity as in FFM. As can be seen in the Table \ref{tab:params-comparison}, in practice the number of parameters are dominated by the term $mk$ and the term $\frac{n(n-1)}{2}k^2$ adds only a few additional parameters because both $n \ll m$ and $k \ll m$. 

\subsubsection{Model complexity}
Table \ref{tab:params-comparison} shows a comparison of the number of parameters used by various models. It also describes why FwFM and FEFM have only slightly higher model complexity as FM, while FFM have a very high model complexity.

\begin{table}[h!]
	\begin{center}
		\begin{tabular}{|c|c|}
			\hline
			\textbf{Model} & \textbf{No. of parameters}  \\ \hline \hline
			LR & $m + 1$  \\ \hline
			FMs &  $m + mk + 1$  \\ \hline
			FFMs & $m + m(n-1)k + 1$  \\ \hline
			FwFMs & $m + mk +\frac{n(n-1)}{2} + 1$  \\ \hline
			FEFMs &  $m + mk + \frac{n(n-1)}{2}k^2 + 1$  \\ \hline
		\end{tabular}
	\end{center}
	\caption{A comparison of the number of parameters used by various models. Here $m$ is the number of unique features, $n$ is the number of unique fields, and $k$ is the dimension of the embedding vectors used to represent features. Since in practice $k << m $ and $n << m$, $mk$ and $m(n-1)k$ terms dominate the number of parameters. FM, FwFM, and FEFM have roughly the same number of parameters for a given $k$ while FFM has a many folds more number of parameters}
	\label{tab:params-comparison}
\end{table}

\subsubsection{FEFM generalizes other shallow Factorization machines}
In equation~\ref{eq:fefm-modeling}, when $W_{F(i),F(j)} = I$, where $I$ is the identity matrix, FEFM reduces to FM. When $W_{F(i)F(j)} = diag(r_{F(i),F(j)})$, where $diag(a)$ is the diagonal matrix with all entries in the diagonal being $a$ and $r_{F(i),F(j)}$ represents field strength between fields $F_i$ and $F_j$, FEFM reduces to FwFM. Thus FEFM is a powerful algorithm with the ability to generalize FM and FwFM. Since $W_{F(i),F(j)}$ transforms $v_j$ and then a dot product is done between the transformed vector and $v_i$, FEFM also has the ability to implicitly generate field-specific feature vector embeddings as explicitly done by FFM.

\subsubsection{Field pair matrix embeddings and field interaction strength}
\label{subsec:interaction-strength}
In equation~\ref{eq:fefm-modeling}, $W_{F(i),F(j)} \in  \mathbb{R}^{k \times k}$ is a symmetric matrix, where $k$ is the dimension of the feature vector embedding space. By the principal axis theorem, $W_{F(i),F(j)}$ has $k$ real non-increasing eigen values $\lambda_1, \lambda_2, . . ., \lambda_k$ and the corresponding orthonormal eigenbasis $u_1, u_2, . . ., u_k$ for the space $\mathbb{R}^k$ .  For the equation~\ref{eq:fefm-modeling} consider the feature interaction term  $v_i^TW_{F(i),F(j)}v_j$ which models the interaction between the feature $i$ of the field $F(i)$ and the feature $j$ of the field  $F(j)$. The feature embeddings $v_i$ and $v_j$ can be represented as linear combinations of the eigenbasis as $v_i = \sum_{t=1}^{k}b_tu_t $ and $v_j = \sum_{t=1}^{k}c_tu_t$. Thus the feature interaction term can be expressed as shown in the equation~\ref{eq:fieldpairinteraction}

\begin{equation}
\label{eq:fieldpairinteraction}
\begin{split}
v_i^TW_{F(i),F(j)}v_j = [\sum_{t=1}^{k}b_tu_t]^TW_{F(i),F(j)}[\sum_{t=1}^{k}c_tu_t]  = \sum_{t=1}^{k} \lambda_t b_t c_t
\end{split}
\end{equation}

where $b_t$ and $c_t$ are the coordinates of the feature emebddings $v_i$ and $v_j$ along the eigenbasis $u_t$. This suggests that the field pair embeddings matrices with eigen values of greater magnitude imply stronger field pair interactions. Thus we define the field pair interaction strength between fields $F(i)$ and $F(j)$ as given in Equation~\ref{eq:fieldpairinteractionstregnth}.

\begin{equation}
\label{eq:fieldpairinteractionstregnth}
\begin{split}
Strength_{F(i),F(j)} = \sqrt{\sum_{t=1}^{k}\lambda_t^2}
\end{split}
\end{equation}

where $\lambda_t$ is the $t^{th}$ eigen value of the field pair embedding matrix $W_{F(i),F(j)}$. Further, the vector embeddings of	 strongly interacting features of a field pair tend to get aligned along the eigenbasis corresponding to the strongest eigen values of the field pair matrix embedding. In section~\ref{subsec:studyfieldinteraction}, we study field pair strength properties on real world datasets.

\subsection{Deep Field-Embedded Factorization Machines}
In this section, we describe how we use FEFM to generate feature interaction embeddings which are adapted into a deep learning paradigm to yield Deep Field-Embedded Factorization Machine (DeepFEFM). The modeling of the logit function for DeepFEFM is shown in Equation~\ref{eq:deepfefmm-modeling}.

\begin{equation}
\label{eq:deepfefmm-modeling}
\begin{split}
\phi(\theta, x) = \phi_{DeepFEFM}((v,w,W, W^{deep}), x) =  \phi_{FEFM}((w, v, W), x)+ \phi_{DNN}((W^{deep}, w^{logit}),  v_{tr})
\end{split}
\end{equation}

Here $\phi_{FEFM}((w, v, W), x)$ follows the same modeling as described in Equation~\ref{eq:fefm-modeling} and $\phi_{DNN}((W^{deep}, w^{logit}),  v_{tr})$ follows the usual modeling of a feed forward DNN used by other deep learning models for CTR prediction. Note the difference that for DeepFEFM in Equation~\ref{eq:deepfefmm-modeling} the input to the DNN is $v_{tr}$, a concatenation of the FEFM interaction embeddings generated by the FEFM interaction module and the input feature vector embeddings. The input layer to the DNN is defined as:

\begin{equation}
\label{eq:deepfefmm-input}
v_{tr} =  concat(\{v_{fefm} ,v_{active(x, 1)}, v_{active(x, 2)}, . . ., v_{active(x, n)}\})
\end{equation}
where $active(x, F)$ is the feature that is active for the data point $x$ for the field $F$. The function $concat(...)$ concatenates all the vectors to create a single vector. And the FEFM interaction embedding $v_{fefm}$ generated by the FEFM interaction module is defined in Equation~\ref{eq:fefm-vec-embedding}

\begin{equation}
\label{eq:fefm-vec-embedding}
v_{fefm} = concat(\{[v_i^TW_{F(i),F(j)}v_jx_ix_j] \mid (i,j) \in \{1,m\} \times \{1,m\} \land i < j\})
\end{equation}

Another way to think about the component FEFM embeddings of $v_{fefm}$ is in terms of the interaction of each feature with other features of different fields. The $i^{th}$ feature interacts with the features of the other $n-1$ fields via the FEFM interaction layer by the operation $v_i^TW_{F(i),F(j)}v_jx_ix_j$. Thus for each feature an FEFM interaction vector embedding of length $n-1$ is generated. All of the elements of these embeddings are concatenated with deduplication in Equation~\ref{eq:fefm-vec-embedding}. 

\begin{figure}[h]
	\centering
	\includegraphics[width=\linewidth]{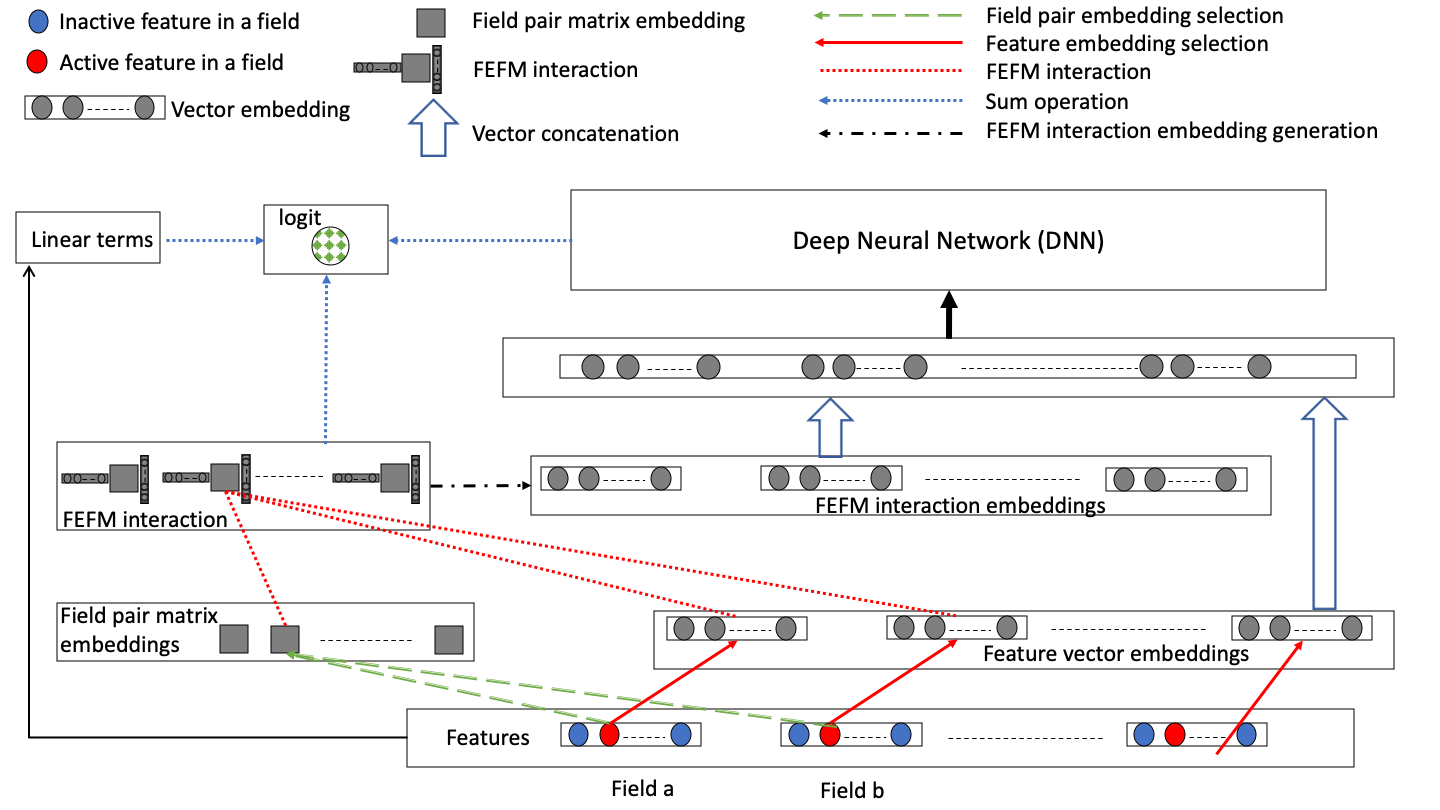}
	\caption{Architecture of Deep Field-Embedded Factorization Machine}
	\label{fig:deepfefmarch}
\end{figure}

The overall architecture of DeepFEFM is shown in Figure~\ref{fig:deepfefmarch}. The ablation studies in Section~\ref{sec:experimentsresults} show that both FEFM interaction embeddings and FEFM logit terms are important for the success of the proposed architecture.

\section{Experiments}
\label{sec:experimentsresults}

In this section, we conduct extensive experiments for CTR prediction tasks to answer the following questions:
\begin{itemize}
	\item \textbf{(Q1)} Over a wide range of hyperparameters, how does our proposed shallow model perform as compared to the state-of-the-art shallow models?
	\item \textbf{(Q2)} Over a wide range of hyperparameters, how does our proposed deep model perform as compared to the state-of-the-art deep models?
	\item \textbf{(Q3)} What is the effect of feature vector embedding dimension on the performance of our proposed models vis-a-vis other state-of-the-art models?
	\item \textbf{(Q4)} How does the validation loss changes with the epochs for our models as compared to the state-of-the-art models?
	\item \textbf{(Q5)}: Which fields and field interactions are important?
	\item \textbf{(Q6)} Ablation studies: How important are various components and design choices of our proposed models?
\end{itemize}

Before addressing these questions we present fundamental details of experimental setup.

\subsection{Datasets}
\label{subsec:datasets}
We use two publicly available datasets called Avazu~\cite{avazu2015} and Criteo~\cite{criteo2014} to compare our proposed models with the existing models. After shuffling,  both the datasets are further divided into train, validation, and test sets. Avazu dataset has only categorical features while  Criteo has both categorical and numerical features. The numerical features in Criteo dataset are discretized at integer levels and thus treated as categorical features. Discretization of numerical features in Factorization Machines is known to yield better performance~\cite{ffm2016}. For both Avazu and Criteo datasets, features with frequency of occurrence less than 20 are filtered out as a noise removal step. For Avazu, the id field is removed and the hour field is transformed to the 0-23 range. For categorical features, if a feature is missing in the test and validation set, then a special feature category of unknown feature is created for all the fields.  All the transformations are fit on the train set and applied to the train, test and validation sets. For discretization of a numerical feature, if a feature is missing in the test and the validation set, its nearest neighbor from the train set is used as the feature. The processed dataset properties are shown in Table~\ref{tab:processed-data-properties}.

\begin{table}[h!]
	\begin{center}
		\begin{tabular}{|c|c|c|c|c|c|}
			\hline
			\textbf{Dataset} & \textbf{$n$} & \textbf{$m$} & \textbf{train : validation : test}\\
			\hline
			Avazu & 22 & 254,644 &25,874,537 : 6,468,635 : 8,085,794  \\ \hline
			Criteo & 39 & 1,040,123 &29,337,994 : 7,334,499 :  9,168,124 \\ \hline		
		\end{tabular}
	\end{center}
	\caption{The number of fields ($n$), features ($m$) and the size of the train, validation, and test tests for both the datasets after preprocessing.}
	\label{tab:processed-data-properties}
\end{table}

\subsection{Implementation and Experimental Setup}
For FFM we use the LIBFFM implementation provided by the authors of FFM~\cite{ffm2016}. All the other models are implemented using Keras~\cite{chollet2015keras}, TensorFlow~\cite{tensorflow2015-whitepaper}, and DeepCTR~\cite{deepctr-library}. For optimization, we use the AdaGrad optimizer~\cite{adagrad2011adaptive}. A batch size of 1024 was used for training. For embeddings at dimensions higher than 32 for Criteo dataset, if the GPU memory is exceeded, a batch size of 512 is used. We discovered that our proposed methods and existing methods show better performance when different L2 regularization is used for the linear parameters, the feature embeddings, and the field-pair embeddings. Thus in our implementation, wherever valid,  we provide support for different L2 regularization for these three parameters. All training and inference was done on Tesla K80 GPUs.

\subsection{Hyperparameter tuning}
To ensure fairness, for all the type of models a wide range of hyperparameters are tried to arrive at their corresponding best hyperparameters. For each type of model and embedding dimension, the hyperparameters yielding the lowest log loss on the validation set are chosen as the best hyperparameters. The corresponding best model is then evaluated on the test set and the performance is reported. Early stopping (minimum delta of 0.000005 and patience of 2) based on the log loss on a validation set was also used during the training of the models~\cite{chollet2015keras}.

\subsection{Comparison with state-of-the-art Shallow Models (Q1)}
\label{sec:comparison-shallow}
For the existing shallow models and the proposed shallow model FEFM, the performance of the best model instances are compared in Table~\ref{tab:performance-comparison-shallow}. For both the datasets, FEFM achieves better performance than the existing state-of-the-art shallow models for both log loss and AUC metrics. For Criteo, FEFM needs the least number of parameters while for Avazu the number of parameters are slightly higher than those of FM and FwFM.

\begin{table*}[t]
	
	\begin{subtable}{\linewidth}
		\begin{center}
			\begin{tabular}{|c|c|c|c|c|}
				\hline
				\textbf{Model} &\textbf{Hyperparameters} & \textbf{AUC} & \textbf{Log loss} & \textbf{\# params} \\ 
				& & & & \textbf{(Mil)} \\
				\hline
				\hline
				LR &  $\eta = 0.05,  \lambda _1 = 1e-7$ &  $0.79416$ & $0.45600$ & $1.0$ \\ \hline
				FM & $\eta = 0.05, \lambda _1 = 1e-6, \lambda _2 = 1e-5,, k = 48$ & $0.80889$  & $0.44269$  & $51.0$ \\ \hline
				FFM &  $\eta = 0.01, \lambda_1 = 1e-7, \lambda_2 = 1e-7, k = 32$  & $0.81018$ & $0.44171$ & $1265.8$ \\ \hline
				FwFM &  $\eta = 0.05, \lambda _1 = 1e-6, \lambda _2 = 1e-5, \lambda _3 = 1e-8, k = 32$ & $0.81024$ & $0.44148$ & $34.3$\\ \hline
				FEFM  &  $\eta = 0.05, \lambda _1 = 1e-6, \lambda _2 = 1e-5, \lambda _3 = 1e-7, k = 16$ & $\textbf{0.81122}$ & $\textbf{0.44053}$ & $17.9$ \\
				(proposed)	& & & & \\ \hline
			\end{tabular}\
			\label{tab:shallow:criteo:comparison}
			\caption{Criteo}
		\end{center}
	\end{subtable}
	
	\begin{subtable}{\linewidth}
		\begin{center}
			\begin{tabular}{|c|c|c|c|c|}
				\hline
				\textbf{Model} & \textbf{Hyperparameters} & \textbf{AUC} & \textbf{Log loss} & \textbf{\# params} \\ 
				& & & & \textbf{(Mil)} \\
				\hline
				\hline
				LR &  $\eta = 0.05,  \lambda _1 = 1e-8$ &  $0.76056$ & $0.38993$ & $0.2$ \\ \hline
				FM &  $\eta = 0.05, \lambda _1 = 1e-6, \lambda _2 = 1e-6, k = 96$ & $0.77474$  & $0.38249$  & $24.7$ \\ \hline
				FFM &  $\eta = 0.6, \lambda_1 = 1e-5, \lambda_2 = 1e-5, k = 16$  & $0.77614$ & $0.38214$ & $85.8$ \\ \hline
				FwFM &  $\eta = 0.1, \lambda _1 = 1e-5, \lambda _2 = 1e-7, \lambda _3 = 1e-8, k = 80$ & $0.77680$ & $0.38139$ & $20.6$ \\ \hline
				FEFM &  $\eta = 0.02, \lambda _1 = 1e-6, \lambda _2 = 1e-6, \lambda _3 = 1e-8, k = 96$ & $\textbf{0.77704}$ & $\textbf{0.38107}$ & $26.8$ \\
				(proposed)	& & & & \\ \hline
			\end{tabular}
			\caption{Avazu}
			\label{tab:shallow:avazu:comparison}
		\end{center}
	\end{subtable}
	
	\caption{A comparison of the prediction performance of FEFM  with the existing state-of-the-art shallow models on the test sets. The best hyperparameters and the total number of parameters for each of the model types are also mentioned. $\lambda _1$, $\lambda _2$, and  $\lambda _3$ are the L2 regularization strengths used for the linear parameters, feature embeddings, and field-pair embeddings; $\eta$ is the learning rate, and $k$ is the size of the feature embeddings.}
	\label{tab:performance-comparison-shallow}
\end{table*}

\subsection{Comparison with state-of-the-art Deep Models (Q2)}
\label{sec:comparison-deep}
For the existing deep models and the proposed deep model DeepFEFM, the performance of the best model instances are compared in Table~\ref{tab:performance-comparison-deep}. For both the datasets, DeepFEFM achieves better performance than the existing state-of-the-art shallow models for both log loss and AUC metrics. FiBiNet is the second in performance and it needs significantly higher number of parameters compared to our proposed DeepFEFM. Another interesting observation is that for Criteo, all the deep models outperform all the shallow models but for Avazu only FiBiNet and DeepFEFM outperform all the shallow models. Interestingly, for Avazu, shallow models FwFM and FEFM (Table~\ref{tab:performance-comparison-shallow}) are able to outperform many other existing deep learning models. This also suggests that deep learning still has a long way to reach the performance level it has reached in images and text models, where deep learning completely outshine shallow learning.

\begin{table*}[t]
	
	\begin{subtable}{\linewidth}
		\begin{center}
			\begin{tabular}{|c|c|c|c|c|}
				\hline
				\textbf{Model} &\textbf{Hyperparameters} & \textbf{AUC} & \textbf{Log loss} & \textbf{\# params} \\ 
				& & & & \textbf{(Mil)} \\
				\hline
				\hline
				DeepFM  &   $\eta = 0.01, \lambda _1 = 1e-6, \lambda _2 = 1e-5,  k = 64$ & $0.81235$ & $0.43933$ & $72.3$\\
				& $\lambda _{deep} = 1e-7, L = 3, dropout = 0.2$ & & & \\ \hline
				xDeepFM & $\eta = 0.01, \lambda _1 = 1e-7 , \lambda _2 = 1e-5, \lambda _c = 1e-6 , k = 16$ & $0.81214$  & $0.43953$  & $22.4$ \\
				& $\lambda _{deep} = 1e-7, L = 3, dropout = 0.2,$ & & & \\ \hline
				AutoInt+  & $\eta = 0.01, \lambda _2 = 1e-5, k = 32, \lambda _{deep} = 0.0$ & $0.81216$  & $0.43963$  & $35.7$ \\ 
				&$ L = 2, dropout = 0.0, H=2, k_a = 64,  L_a= 1 $& & & \\ \hline
				FiBiNet & $\eta = 0.01, \lambda _1 = 1e-6, \lambda _2 = 1e-6, k = 64$ & $0.81302$  & $0.43871$  & $173.1$ \\ 
				& $\lambda _{deep} = 0.0, L = 3, dropout = 0.5, r = 3$ & & & \\ \hline
				DeepFEFM   & $\eta = 0.01, \lambda _1 = 1e-6, \lambda _2 = 1e-5, \lambda_3 = 1e-7, k = 48$ & $\textbf{0.81405}$  & $\textbf{0.43778}$  & $57.6$ \\
				(proposed)	& $\lambda _{deep} = 1e-7, L = 3, dropout = 0.2$ & & & \\ \hline
			\end{tabular}
			\label{tab:deep:criteo:comparison}
			\caption{Criteo}
		\end{center}
	\end{subtable}
	
	\begin{subtable}{\linewidth}
		\begin{center}
				\begin{tabular}{|c|c|c|c|c|}
				\hline
				\textbf{Model} &\textbf{Hyperparameters} & \textbf{AUC} & \textbf{Log loss} & \textbf{\# params} \\ 
				& & & & \textbf{(Mil)} \\
				\hline
				\hline
				DeepFM  &  $\eta = 0.01, \lambda _1 = 1e-6, \lambda _2 = 1e-6,  k = 64$ & $0.77594$ & $0.38148$ & $20.1$ \\
				& $\lambda _{deep} = 1e-7, L = 3, dropout = 0.2$ & & & \\ \hline
				xDeepFM & $\eta = 0.01, \lambda _1 = 1e-6 , \lambda _2 = 1e-5, \lambda _c = 1e-5 , k = 32$ & $0.77550$  & $0.3816$  & $12.8$ \\
				& $\lambda _{deep} = 1e-7, L = 3, dropout = 0.2,$ & & & \\ \hline
				AutoInt+  & $\eta = 0.01, \lambda _2 = 1e-5, k = 80, \lambda _{deep} = 1e-7$ & $0.77627$  & $0.38153$  & $24.4$ \\ 
				&$ L = 3, dropout = 0.0, H=2, k_a = 80,  L_a= 1 $& & & \\ \hline
				FiBiNet & $\eta = 0.01, \lambda _1 = 1e-7, \lambda _2 = 1e-6, k = 96$ & $0.77706$  & $0.38083$  & $76.6$ \\ 
				& $\lambda _{deep} = 0.0, L = 3, dropout = 0.5, r = 3$ & & & \\ \hline
				DeepFEFM   & $\eta = 0.01, \lambda _1 = 1e-7, \lambda _2 = 1e-6, \lambda_3 = 1e-8, k = 96$ & $\textbf{0.77801}$  & $\textbf{0.38052}$  & $31.4$ \\
				(proposed)	& $\lambda _{deep} = 1e-7, L = 3, dropout = 0.2$ & & & \\ \hline
			\end{tabular}
			\caption{Avazu}
			\label{tab:deep:avazu:comparison}
		\end{center}
	\end{subtable}
	
	\caption{A comparison of the prediction performance of DeepFEFM  with the existing state-of-the-art deep models on the test sets. The best hyperparameters and the total number of parameters for each of the model types are also mentioned. $\lambda _1$, $\lambda _2$, and  $\lambda _3$ are the L2 regularization strengths used for the linear parameters, feature embeddings, and field-pair embeddings; $\eta$ is the learning rate and $k$ is the  size of the feature embeddings; $\lambda _{deep}$ is the L2 regularization strength of the DNN component, $L$ for the number of DNN layers (excluding input layer and logit layer), the width of all the DNN layers is $1024$ for all the models; $\lambda _c$ is the L2 regularization strength for the CIN module of xDeepFM;  $H$ is the number of self-attention heads, $k_a$ is the attention embedding dimension,  $L_a$ is the number of attention layers in AutoInt+; $r$ is the reduction ratio in FiBiNet }
	\label{tab:performance-comparison-deep}
\end{table*}

\subsection{Effect of Embedding Dimension on Performance (Q3)}
\label{sec:comparison-embeddings}

\begin{figure}[!htbp]
	\centering
	\includegraphics[width=\linewidth]{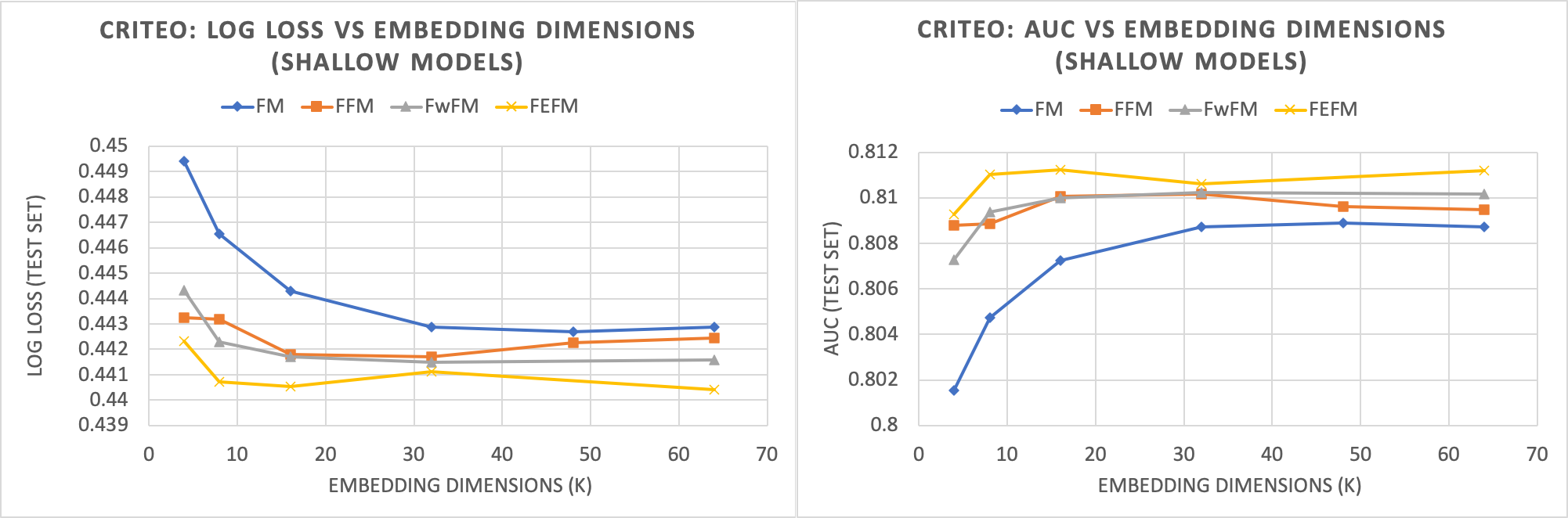}
	\caption{Criteo test set: Log loss and AUC comparison with state-of-the-art shallow models}
	\label{fig:ShallowCriteoLossAUCVSk}
\end{figure}

\begin{figure}[!htbp]
	\centering
	\includegraphics[width=\linewidth]{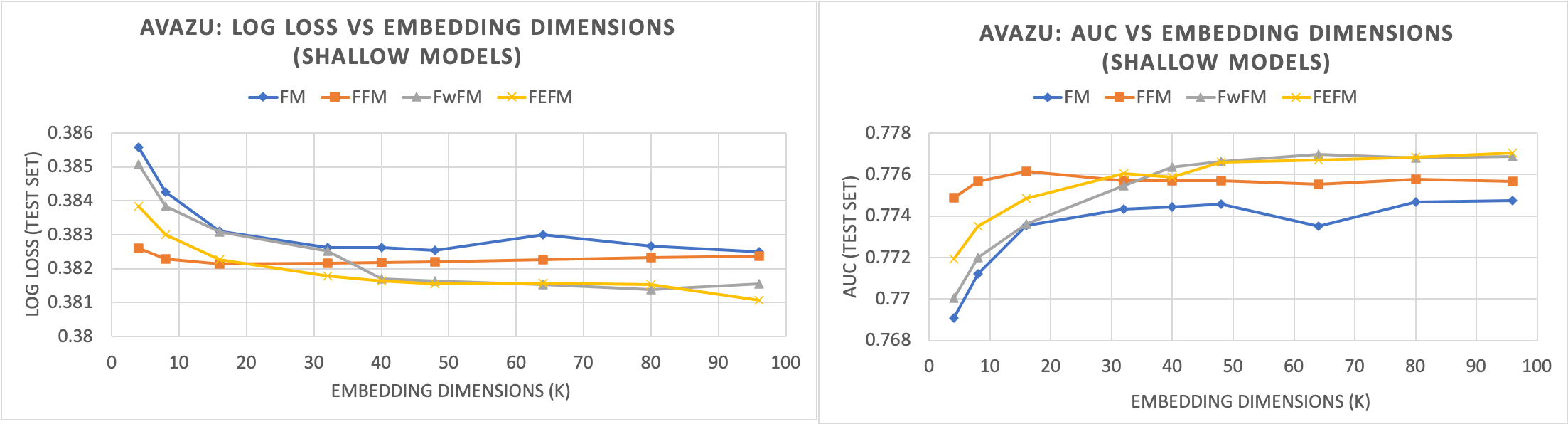}
	\caption{Avazu test set: Log loss and AUC comparison with state-of-the-art shallow models}
	\label{fig:ShallowAvazuLossAUCVSk}
\end{figure}

\begin{figure}[!htbp]
	\centering
	\includegraphics[width=\linewidth]{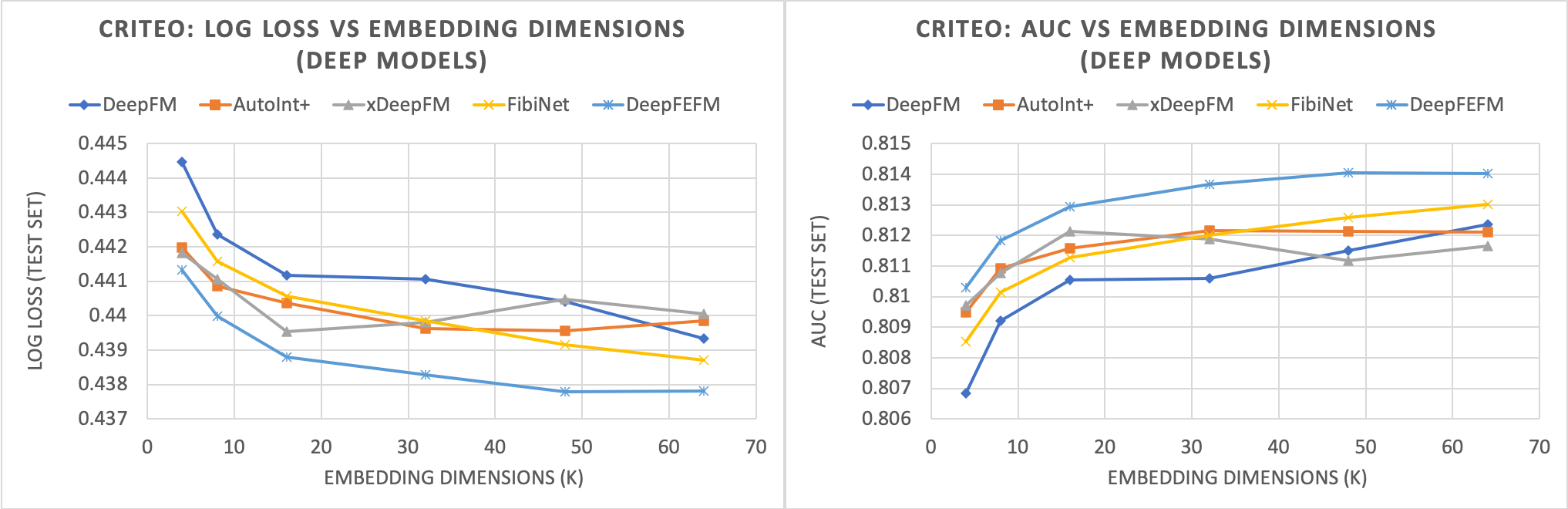}
	\caption{Criteo test set: Log loss and AUC comparison with state-of-the-art deep models}
	\label{fig:DeepCriteoLossAUCVSk}
\end{figure}

\begin{figure}[!htbp]
	\centering
	\includegraphics[width=\linewidth]{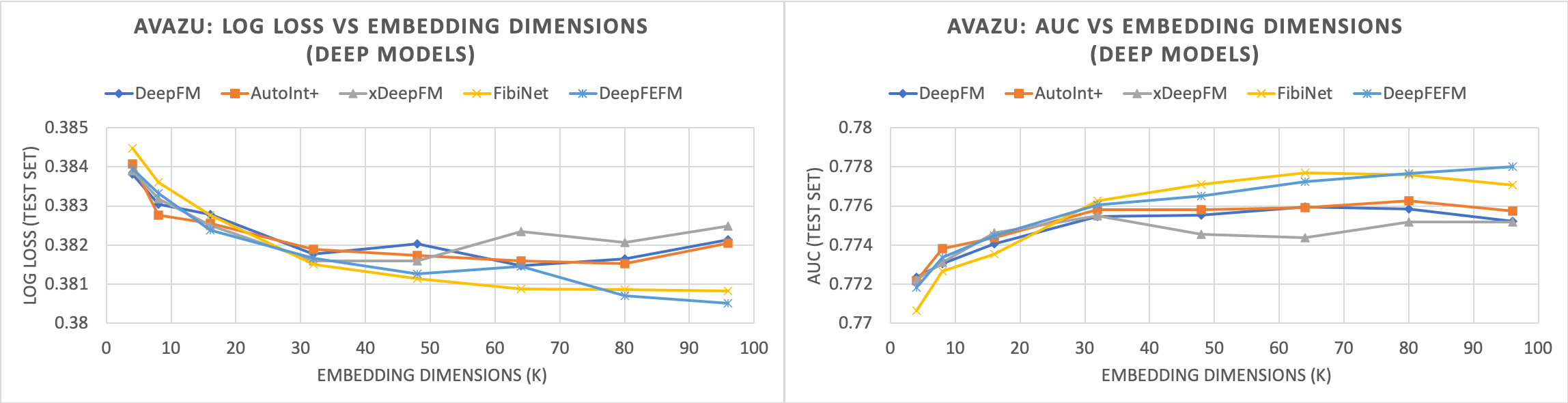}
	\caption{Avazu test set: Log loss and AUC comparison with state-of-the-art deep models}
	\label{fig:DeepAvazuLossAUCVSk}
\end{figure}

Most of the existing deep learning work on CTR models report comparison at a single embedding dimension and don't the compare performance of their models with existing models for different embedding dimensions. However, earlier work on shallow models such as FFM~\cite{ffm2016} does perform a thorough comparison across embedding dimensions. We found that different models behave differently at different dimensions and thus we compare the performance of both shallow (Figure~\ref{fig:ShallowCriteoLossAUCVSk} and ~\ref{fig:ShallowAvazuLossAUCVSk}) and deep models (Figure~\ref{fig:DeepCriteoLossAUCVSk} and ~\ref{fig:DeepAvazuLossAUCVSk}). It is interesting to observe that the behavior changes not only across different model types but also across the datasets. For Criteo, FEFM outperforms all the other shallow models at all dimensions. For Avazu, at lower dimensions FFM outperforms other models but is soon overtaken by both FwFM and FEFM. At the extremities, FEFM outperforms FwFM while for the moderate dimensions, their performance is competitive. For deep learning models on Criteo, DeepFEFM outperforms the other deep models for all dimensions. At lower dimensions, xDeepFM and AutoInt+ are second to DeepFEFM, but xDeepFM and AutoInt+ tend to overfit at higher dimensions. For Avazu, the difference in performance is narrow at lower dimensions and it widens at higher dimensions. At very low dimensions AutoInt+ performs better but overfit at moderate and higher dimensions. FiBiNet performs the best at moderate dimensions but not at the extremities. At lower dimensions, FibiNet is the weakest of all in performance. Also xDeepFM and AutoInt+ encounter a quick drop in performance at higher dimensions while FiBiNet and DeepFEFM continue to perform better at higher dimensions. While for the moderate dimensions, FiBiNet is better, at the extremities, DeepFEFM outperforms FiBiNet.

\subsection{Convergence of training (Q4)}
\begin{figure}[t]
	\centering
	\includegraphics[width=\linewidth]{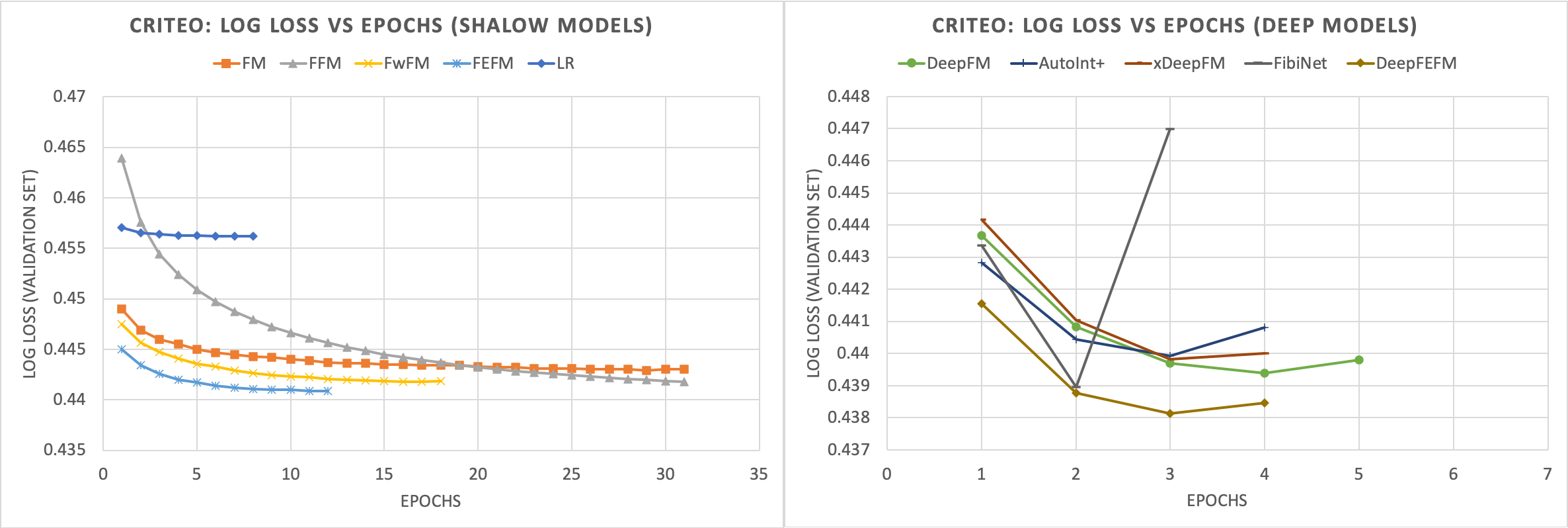}
	\caption{Criteo validation set: Log loss for the shallow and deep models at different epochs.}
	\label{fig:CriteoLossVSepochs}
\end{figure}

\begin{figure}[!htbp]
	\centering
	\includegraphics[width=\linewidth]{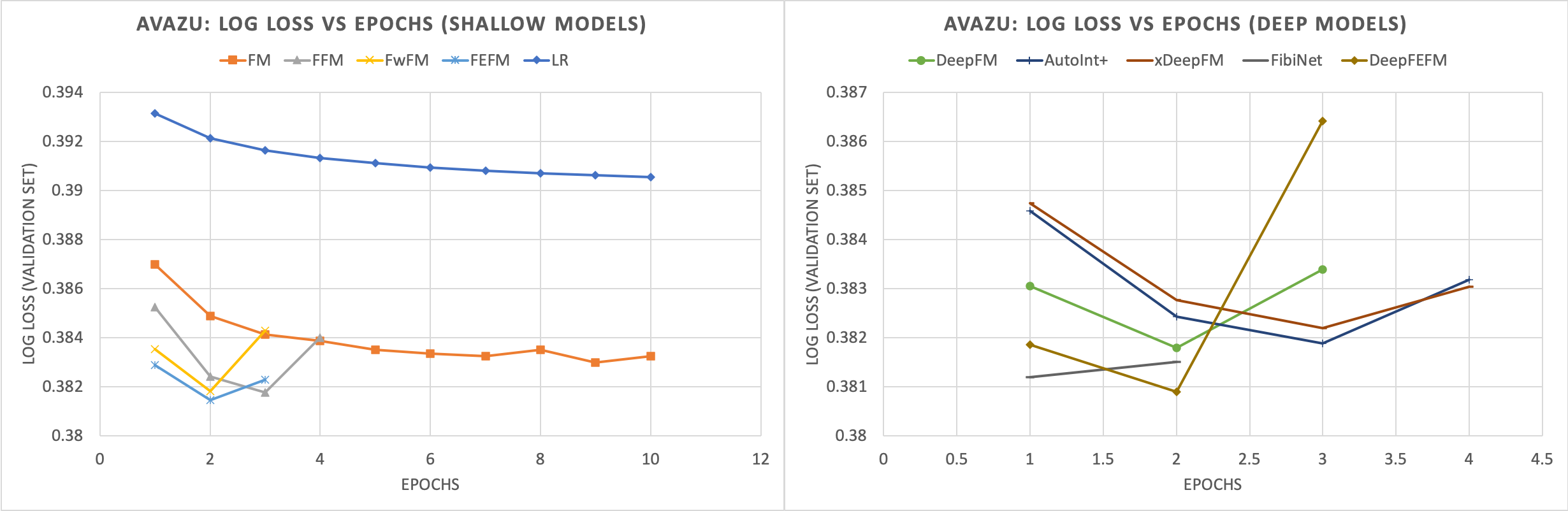}
	\caption{Avazu validation set: Log loss for the shallow and deep models at different epochs.}
	\label{fig:AvazuLossVSepochs}
\end{figure}

Figure~\ref{fig:CriteoLossVSepochs} and ~\ref{fig:AvazuLossVSepochs}  show how the log loss on the validation set changes with each epoch for the best models for each type for both the datasets. We can see that overall for all the models, convergence is faster for Avazu  as compared to Criteo. For shallow FM-variants, FEFM and FwFM converge faster than FFM and FM. Deep models in general converge quickly, with FiBiNet being the fastest to converge, which is expected because FiBiNet uses very high number of parameters compared to the other models.

\subsection{Study of field interaction strength (Q5)}
\label{subsec:studyfieldinteraction}

\begin{figure}[!htbp]
	\centering
	\begin{subfigure}{0.5\textwidth}
		\centering
		\includegraphics[width=0.65\textwidth]{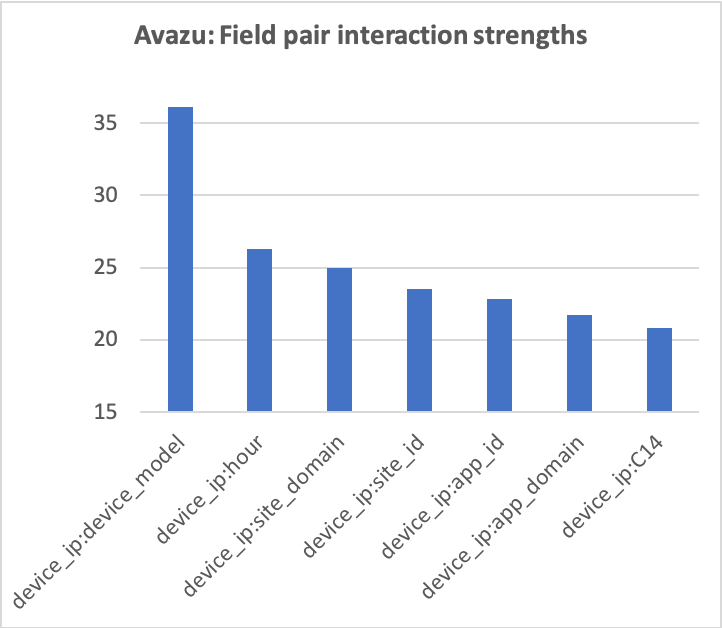}
		\caption{Avazu: Top 7 field pairs reverse sorted by their interaction strengths}
		\label{fig:avazufield:pairs}
		\label{fig:inclu}
	\end{subfigure}%
	\begin{subfigure}{0.5\textwidth}
		\includegraphics[width=0.9\textwidth]{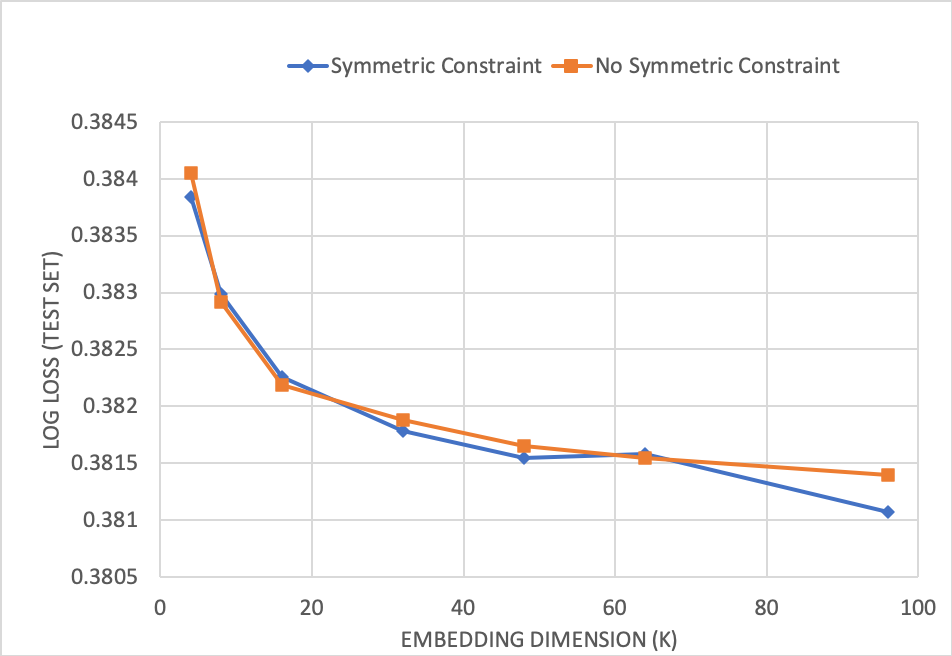}
			\caption{Log loss on Avazu test dataset for symmetric vs non-symmetric matrix embeddings in FEFM}
		\label{fig:avazu:symvsnosym}
	\end{subfigure}
	\caption{Avazu: Field pair strengths and the effect of symmetric constraint on matrix embeddings}
	\label{fig:manmade}
\end{figure}

In Section~\ref{subsec:interaction-strength}, we discussed how FEFM provides insights into the interaction strengths of the field pairs via eigen values of the field pair embedding matrices. The bar charts in Figure~\ref{fig:avazufield:pairs} show the top 7 field pairs arranged in decreasing order of their interaction strengths as per Equation~\ref{eq:fieldpairinteractionstregnth}. For Avazu dataset, it can be found that the top-most field pairs include interaction between ``device\_ip" and other fields such as ``device\_model", ``hour", ``site\_domain", etc. This also makes intuitive sense since ``device\_ip" is a very important field for CTR prediction. Since the field names are completely anonymized for Criteo dataset, we have not presented the plots. We found that the interaction between the anonymized field ``C24" and multiple other fields is very important.

\begin{table}[!htbp]
	\begin{center}
		\begin{tabular}{|c|c|c|c|c|}
			\hline
			\textbf{Model} & \textbf{ Criteo AUC} & \textbf{Criteo LogLoss} & \textbf{Avazu AUC} & \textbf{Avazu LogLoss} \\
			\hline
			 No Ablation & 0.$81405$ & $0.43778$ & $0.77801$ & $0.38052$ \\ \hline
			 Ablation1 & $0.81381$ & $0.43795$ & $0.77689$ & $0.38113$ \\ \hline	
			 Ablation2 & $0.81386$ & $0.43793$ & $0.77768$ & $0.38078$ \\ \hline
			 Ablation3 & $0.81363$ & $0.43832$ & $0.77814$ & $0.3806$ \\ \hline	
			 Ablation4 & $0.81259$ & $0.43919$ & $0.77656$ & $0.38121$ \\ \hline
		\end{tabular}
	\end{center}
	\caption{Comparison of the prediction performance of DeepFEFM with ablation of its various components. 	{\textbf{Ablation1}}: removed the FEFM logit terms from the final logit, {\textbf{Ablation2}}: removed the linear terms from the final logit, {\textbf{Ablation3}}: removed the feature vector embeddings from the DNN input, {\textbf{Ablation4}}: removed the FEFM interaction embeddings from the DNN input.}
	\label{tab:ablation-studies}
\end{table}

\subsection{Ablation Studies (Q6)}
We ablate different components of our models to answer the following specific questions:

\begin{itemize}
	\item{\textbf{What is the effect of the symmetric constraint on field pair matrix embeddings?}}
	In Section~\ref{subsec:interaction-strength} and Section~\ref{subsec:studyfieldinteraction} we discussed the rationale and field interaction insights on public datasets  in regard to the symmetric property of the field pair matrix embeddings. In Figure~\ref{fig:avazu:symvsnosym} we compare the performance of FEFM with and without the symmetric constraint on the field pair matrix embeddings. We see that at lower embedding dimensions, the performance is similar for both the cases. At higher dimensions, the symmetric constraint gives slightly better performance, suggesting a mild regularizing effect when the number of parameters are high. In either case, it is desirable to have the symmetric constraint for its mathematical properties that explain the important fields and their interactions.
	
	\item{\textbf{Which components of the DeepFEFM architecture are the most important?}}
	We compare the performance of different ablated architectures of DeepFEFM in Table~\ref{tab:ablation-studies}. Removal of FEFM interaction embeddings from the proposed DeepFEFM architecture leads to the biggest drop in the performance for both the datasets. Removal of other layers have a minor effect on Criteo but for Avazu, removal of FEFM logit terms leads to the second biggest drop in performance. This leads us to conclude that FEFM interaction embeddings and FEFM logit terms are the two most important components of our architectures. These are also our novel contributions to the CTR prediction models.
	
\end{itemize}

\section{Conclusion}
\label{sec:conclusion}
In this paper, we first proposed a shallow model called Field-Embedded Factorization Machine (FEFM) for click-through rate prediction. FEFM is a novel variant of Factorization Machines (FM) that leverages field-specificity in feature interaction by introducing learnable symmetric matrix embeddings for each field pair. The eigen values of the field pair embeddings represent interaction strengths between the  fields. We then extended the shallow model to a deep learning model DeepFEFM using the FEFM interaction embeddings. By conducting extensive experiments over a wide range of hyperparameters, we showed that both our proposed models consistently outperform existing state-of-the-art models in their respective shallow and deep categories. By conducting ablation studies, we also showed that FEFM logit terms and FEFM interaction embeddings are critical to the boost in prediction performance.

\section*{Acknowledgment}
We would like to thank Abhishek Veeraraghavan, Kunal Kumar Jain, Priyanka Patel, Tathagata Sengupta, and Umang Moorarka for their valuable feedback for the successful completion of this work.

\bibliographystyle{unsrt}

\end{document}